\documentclass[sigconf]{acmart}

\usepackage{subcaption}
\usepackage{xcolor,soul}
\usepackage{amsfonts}
\usepackage{listings}
\newcommand{\true}{\texttt{TRUE}}

\newtheorem{claim}{Claim}[section]
\newtheorem{observation}{Observation}[claim]

\copyrightyear{2023}
\acmYear{2023}
\setcopyright{rightsretained}
\acmConference[HOPC '23] {Proceedings of the 2023 ACM Workshop on Highlights of Parallel Computing }{June 16--19, 2023}{Orlando, FL, USA.}
\acmBooktitle{Proceedings of the 2023 ACM Workshop on Highlights of Parallel Computing (HOPC '23), June 16--19, 2023, Orlando, FL, USA}
\acmPrice{}
\acmISBN{979-8-4007-0218-1/23/06}
\acmDOI{10.1145/XXXXXX.XXXXXX}

\begin{document}

\title{Empirical Challenge for NC Theory}

\author{Ananth Hari}
\email{ahari1@umd.edu}
\orcid{0000-0001-8174-5335}
\affiliation{%
  \institution{University of Maryland}
  \city{College Park}
  \state{Maryland}
  \country{USA}
  \postcode{20742}
}

\author{Uzi Vishkin}
\email{vishkin@umd.edu}
\orcid{0000-0003-0713-2674}
\affiliation{%
  \institution{University of Maryland}
  \city{College Park}
  \state{Maryland}
  \country{USA}
  \postcode{20742}
}

\renewcommand{\shortauthors}{Ananth Hari \& Uzi Vishkin}

\begin{abstract}
  Horn-satisfiability or Horn-SAT is the problem of deciding whether a satisfying assignment exists for a Horn formula, a conjunction of clauses each with at most one positive literal (also known as Horn clauses). It is a well-known \textbf{P}-complete problem, which implies that unless $\textbf{P} = \textbf{NC}$, it is a hard problem to parallelize. In this paper, we empirically show that, under a known simple random model for generating the Horn formula, the ratio of hard-to-parallelize instances (closer to the worst-case behavior) is infinitesimally small. We show that the depth of a parallel algorithm for Horn-SAT is polylogarithmic on average, for almost all instances, while keeping the work linear. This challenges theoreticians and programmers to look beyond worst-case analysis and come up with practical algorithms coupled with respective performance guarantees.
\end{abstract}




\maketitle

\section{Introduction}
A Horn clause is a disjunction of boolean literals of which at most one is positive and a CNF Horn formula (or more succinctly, Horn formula) is a conjunction of Horn clauses. Horn-satisfiability (or Horn-SAT) is the problem of deciding whether a truth assignment exists for all variables that satisfies the formula and if it does, output the assignment. 

Horn-SAT is \textbf{P}-complete (as noted in \citet{dowling1984linear}). This makes it unlikely that there will be an \textbf{NC} (``Nick's Class'', in honor of Nick Pippenger) algorithm for it, namely an algorithm that runs in polylogarithmic (poly-log) time using a polynomial number of processors. An \textbf{NC}-algorithm for a \textbf{P}-complete problem would immediately imply an \textbf{NC}-algorithm for any problem in the class \textbf{P} (in other words, any problem that can be solved using a serial polynomial time algorithm), which is considered unlikely by complexity theorists. So, linear speedup not reaching poly-log time is the best one can hope for in the worst case. The exact upper bound in this work is $O(n/p + h \log n)$ for an input of $n$ literals on a $p$-processor Arbitrary CRCW PRAM. Every iteration of the parallel algorithm seeks to concurrently commit the truth value of as many variables as possible, based on truth values committed in prior iterations. The parameter $h$ is the length of the longest chain of such commit iterations, given the input formula (by rather limited analogy to BFS). This means that the best parallel time (known also as ``depth'') that the parallel algorithm can provide would be $O(h \log n)$, and it will need $n/(h \log n)$ processors to achieve that. While none of the above is surprising, the dominant logarithmic empirical behavior of $h$ offers a perhaps first dramatic contrast to the common wisdom that \textbf{P}-completeness is ``a nightmare for parallel processing'', as expressed in chapter 9 of \citet{mehlhorn2008algorithms}.

The rest of the paper is organized as follows: Section~\ref{sec:prelim} gives a brief introduction on the exact problem formulation, its inputs and outputs. Section~\ref{sec:GP} provides the greedy parallel algorithm to solve a general Horn-SAT formula. Section~\ref{sec:rand} gives the random Horn-SAT model on which we analyze the performance of a slightly modified greedy parallel algorithm. Finally, Section~\ref{sec:emp} puts forward our main claim (along with empirical evidence) of apparent low depth of the parallel algorithm on random Horn formulae.

This paper serves as an extension to our previously published work: \citet{vishkin2022beyond}.

\section{Preliminaries}\label{sec:prelim}
We consider the 3CNF Horn-SAT, i.e., Horn formulae with at most three literals per clause, of which at most one is positive. Appendix~\ref{app:3cnf} gives a linear work reduction from a general Horn formula to a 3CNF Horn formula. The details of problem formulation including the inputs and outputs are detailed in Appendix~\ref{app:prob}, which are taken from \citet{vishkin2022beyond}.

\section{The Greedy Parallel Algorithm}\label{sec:GP}
Similar to \citet{dowling1984linear}, we assume that Horn formulae are in a ``reduced'' form, i.e., that there are no duplicate clauses and no duplicate literals within clauses.

The greedy parallel (GP) algorithm for Horn-SAT was introduced in \citet{vishkin2022beyond}. The pseudocode is presented in Appendix~\ref{app:gp} for the sake of completeness. Section 3 of \citet{vishkin2022beyond}
lays out arguments for correctness and details on implementation leading to $O(n/p + h\log n)$ time and $O(n)$ work complexities.

We use a slightly modified version of the GP algorithm, namely the Parallel Positive Unit Resolution (PPUR) algorithm, that lends itself to more effective evaluation of the parallel algorithm on random Horn formulae. The PPUR algorithm uses resolution of positive unit clauses only, instead of resolving using both negative and positive unit clauses as done by the GP algorithm. More information about the PPUR algorithm is present in Appendix~\ref{app:ppur}.

\section{The Random Horn-SAT model}\label{sec:rand}
The complexity analysis of the algorithms suggests the question: what values for $h$ (the number of parallel rounds of main loop of the PPUR algorithm) can we expect?

Personal communication with three constraint satisfaction experts suggested that the model that \citet{moore2007continuous} discussed for random Horn-SAT formulae is probably the most relevant in the literature. We focus our attention on the random 1-3-Horn-SAT model, the random model corresponding to 3CNF Horn-SAT. For positive two real number $d_1 < 1$ and $d_3$, let the random 1-3-Horn-SAT formula $\mathcal{H}(n,d_1,0,d_3)$ be the conjunction of: (a) a single negative literal $x_1$, (b) $d_1n$ positive literals chosen uniformly without replacement from the variables $x_2, \ldots, x_n$, and (c) $d_3n$ Horn clauses chosen uniformly with replacement from the $\frac{n(n-1)(n-2)}{2}$ possible Horn clauses with 3 variables where one literal is positive. The ``0'' included in the parameter list indicates that there are no initial Horn clauses of length 2 in the input formula.

\citet{moore2007continuous} used this model to study phase transition in probability of satisfiability of the serial algorithm (PUR) on random 1-3-Horn-SAT. They found that, within the domains of $(d_1,d_3)$ values, the probability of satisfiability of the Horn formula $\mathcal{H}(n,d_1,0,d_3)$:
\begin{enumerate}
    \item Is continuous for $d_3 < 2$, for any value of $d_1$ in its domain.
    \item Is discontinuous at a specific tuple $(d_1^*,d_3)$, for each $d_3 \ge 2$. The value of $d_1^*$ as a function of $d_3$ is given in Appendix~\ref{app:tran}.
\end{enumerate}

\section{Empirical depth of the PPUR algorithm on random 1-3-Horn-SAT}\label{sec:emp}

We make the following claim on the expected number of rounds ($h$) taken by the PPUR algorithm to converge on a random 1-3-Horn-SAT instance $\mathcal{H}$, based on empirical evidence. The process we used to empirically compute $h$ for any given initial values of $d_1$ and $d_3$ is given in Appendix~\ref{app:h}.

\begin{claim}\label{cl:1}
  For every random 1-3-Horn-SAT formula $\mathcal{H}(n,d_1,0,d_3)$ where the probability of satisfiability is continuous, the average growth of $h$ (the number of rounds of the PPUR algorithm) is proportional to $\log n$. 
\end{claim}

The above claim is a result of observing the behavior of $h$ for three cases: (i) when $d_3 < 2$, (ii) when $d_3 \ge 2$ and $d_1 \ne d_1^*$, and (iii) when $d_3 \ge 2$ and $d_1 = d_1^*$.

\begin{observation}\label{obs:1}
  For $d_3 < 2$ and for $d_1$ in its domain, the average growth of $h$ is proportional to $\log n$.
\end{observation}

\begin{figure}[ht]
    \centering
    \includegraphics[width=\columnwidth]{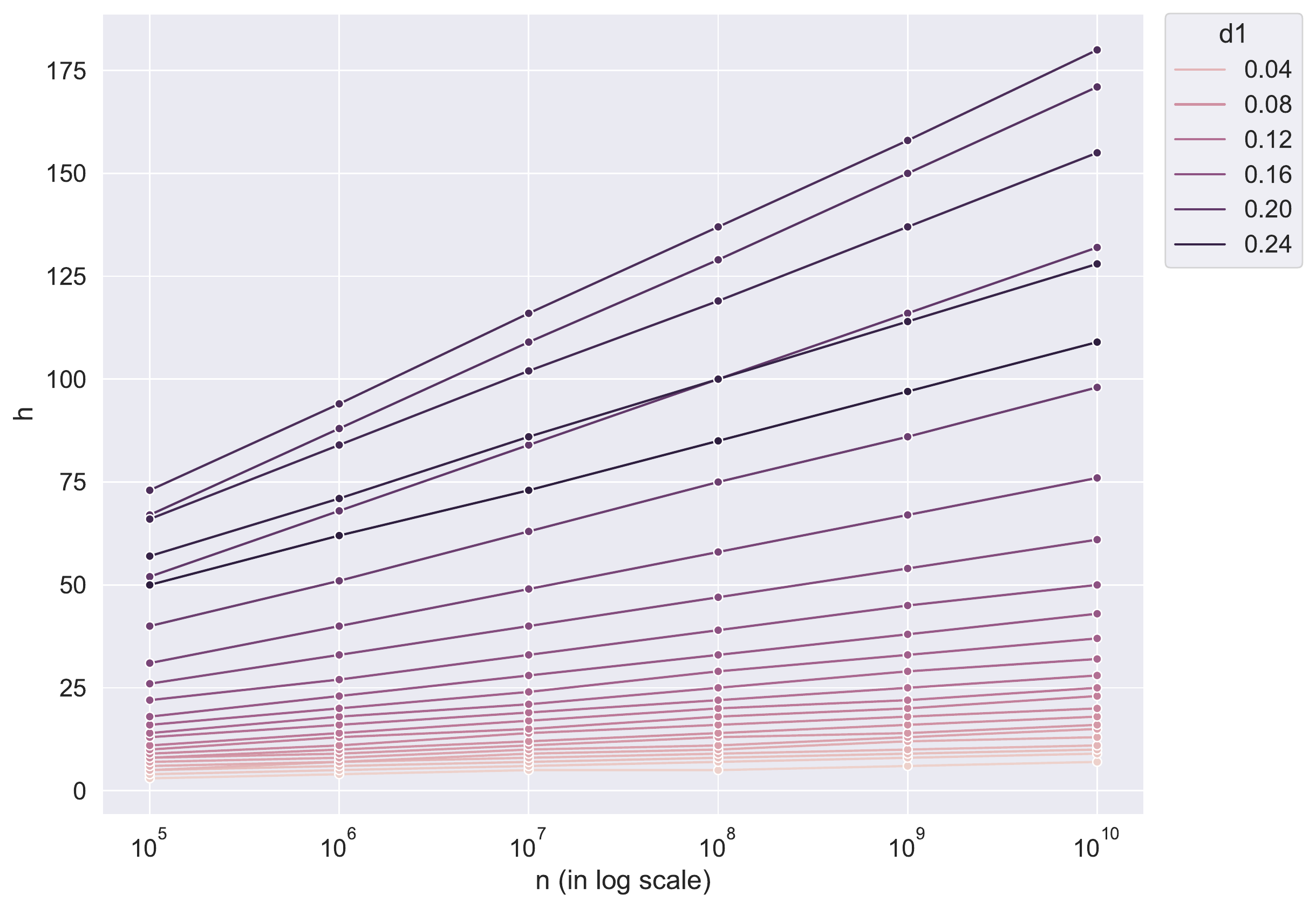}
    \caption{$n$ (in log scale) vs. $h$ for a random 1-3-Horn-SAT formula where $d_3 = 1.8$, for various values of $d_1$. The full legend is not printed to reduce visual clutter. The hue darkens as $d_1$ increases.}
    \label{fig:d3_l_2}
    \Description{A standard line plot is drawn for increasing values of a parameter $d_1$. X-axis is the number of variables in the Horn formula in log scale and Y-axis is the expected number of rounds needed for the PPUR algorithm to converge. All lines in the graph are straight lines with varying slopes and do not intersect the origin.}
\end{figure}

Figure~\ref{fig:d3_l_2} illustrates Observation~\ref{obs:1}, as $h$ appears to grow linearly as $n$ grows exponentially.

\begin{observation}\label{obs:2}
  For $d_3 \ge 2$ and for $|d_1 - d_1^*| \ge \epsilon$, where $\epsilon \in (0, d_1^*)$, the average growth of $h$ is proportional to $\log n$.
\end{observation}

\begin{figure}[ht]
    \centering
    \begin{subfigure}[b]{\columnwidth}
         \centering
         \includegraphics[width=\columnwidth]{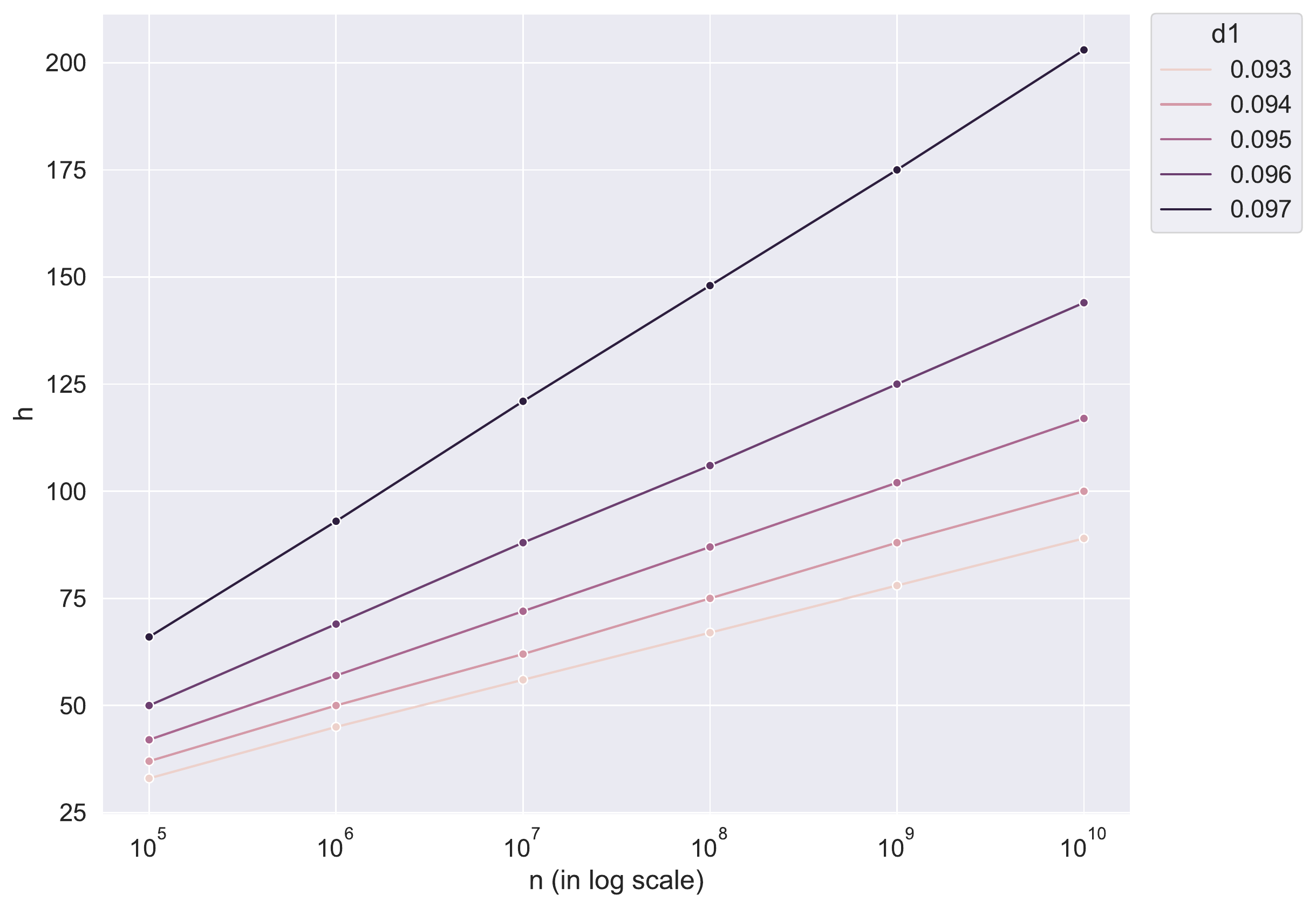}
         \caption{$d_1$ is at most $d_1^* - \epsilon$. Here, $\epsilon$ is chosen to be $0.001$.}
     \end{subfigure}
     \vfill
     \begin{subfigure}[b]{\columnwidth}
         \centering
         \includegraphics[width=\columnwidth]{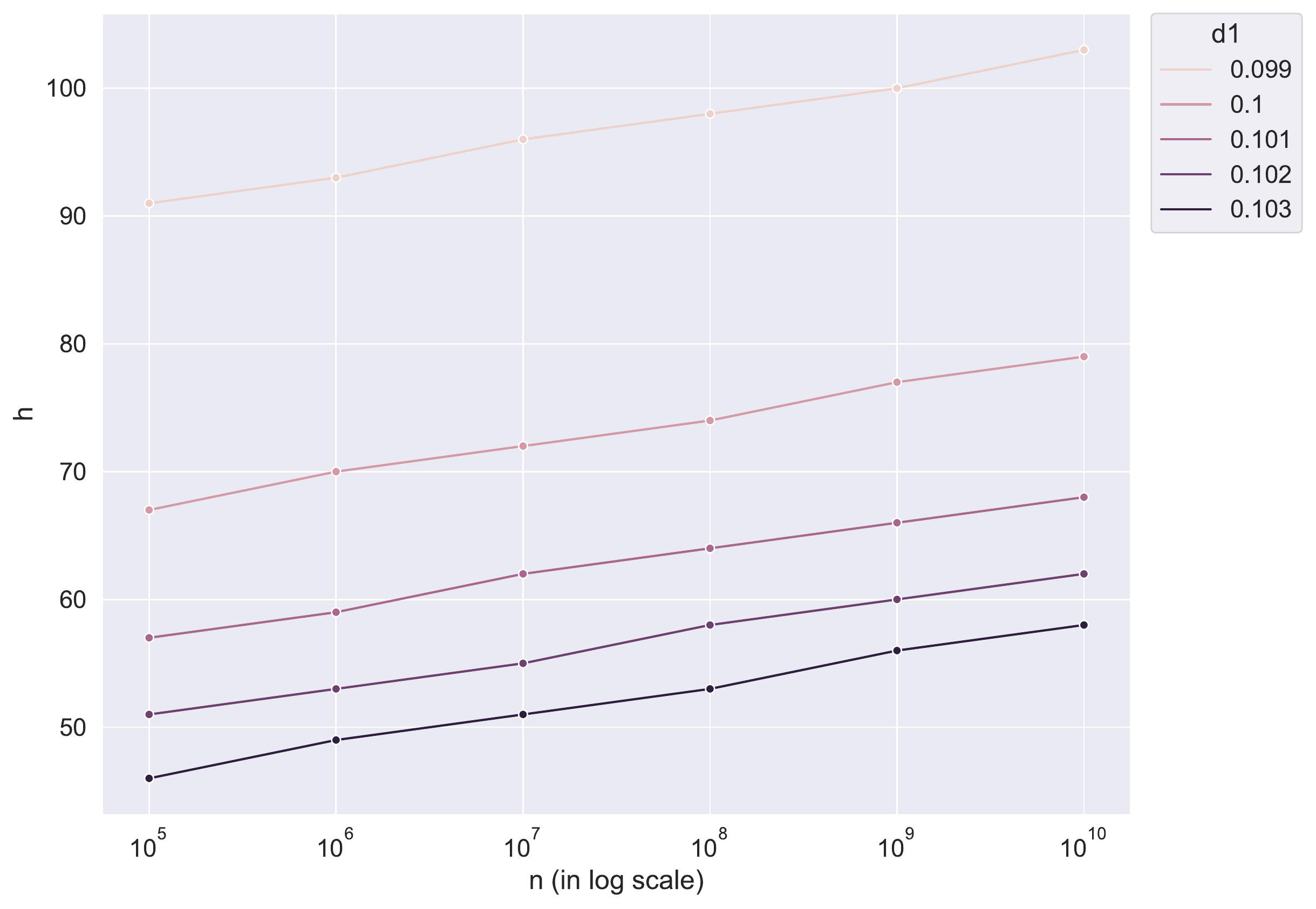}
         \caption{$d_1$ is at least $d_1^* + \epsilon$. Here, $\epsilon$ is chosen to be $0.001$.}
     \end{subfigure}
        \caption{$n$ (in log scale) vs. $h$ for a random 1-3-Horn-SAT formula where $d_3 = 3.0$, for various values of $d_1$. Let $d_1^*$ be the critical value of $d_1$ whose meaning and value are detailed in Appendix~\ref{app:tran}. Here, $d_1^* \approx 0.098$. $h$ increases as $d_1$ gets closer to $d_1^*$ (from either direction). $d_1$ values in the legend are truncated to three significant digits.}
        \label{fig:d3_ge_2}
        \Description{The figure contains two subfigures which show similar trends in data. Each panel has four non-intersecting straight lines with hue of each line going from light to dark in the bottom panel and the opposite in the top panel.}
\end{figure}

Figure~\ref{fig:d3_ge_2} illustrates Observation~\ref{obs:2}, as $h$ appears to grow linearly as $n$ grows exponentially.

\begin{observation}\label{obs:3}
  For $d_3 \ge 2$ and for $d_1 = d_1^*$, the average growth of $h$ is proportional to $n$.
\end{observation}

\begin{figure}[ht]
    \centering
    \includegraphics[width=\columnwidth]{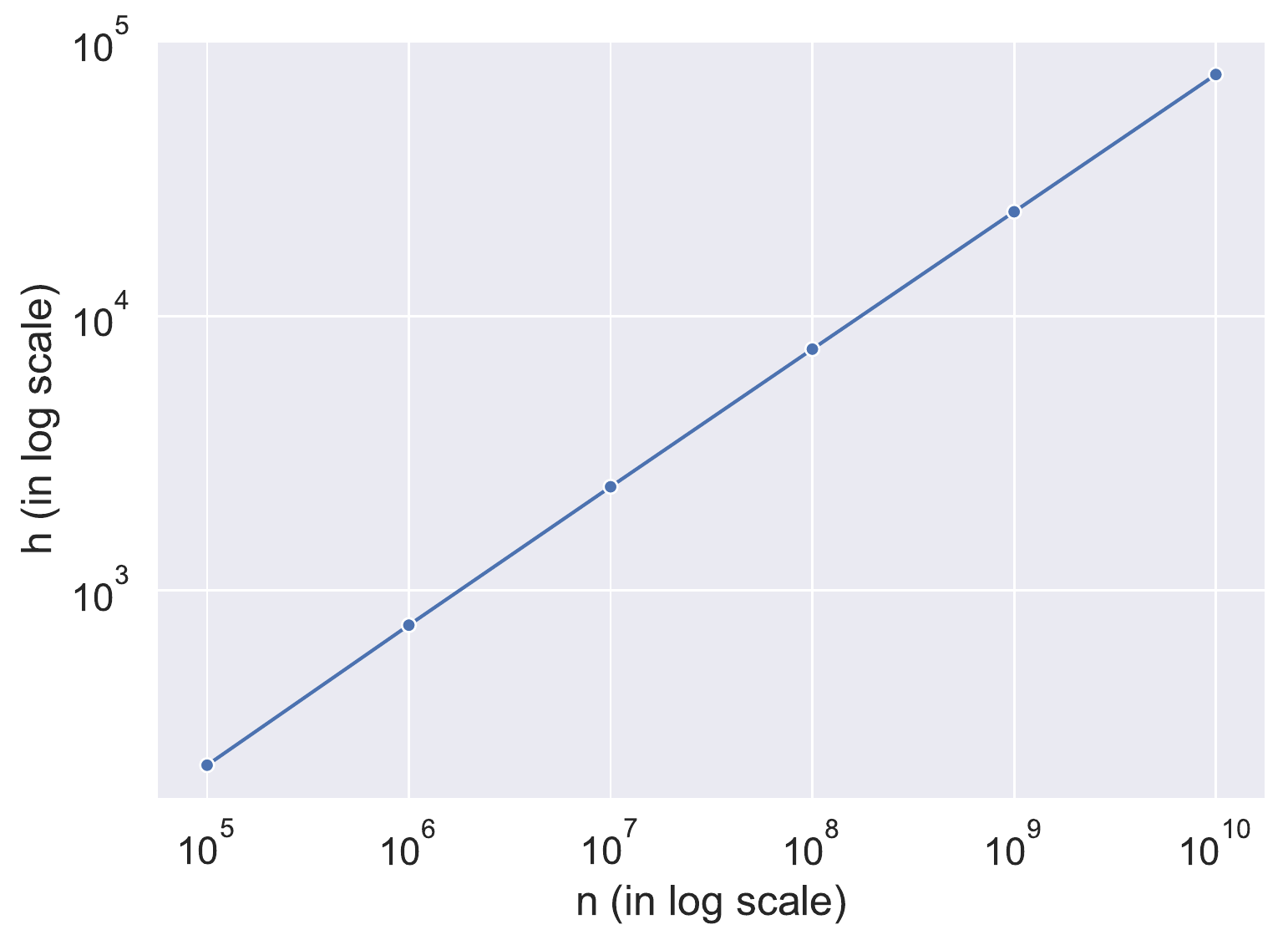}
    \caption{$\log n$ (in log scale) vs. $ \log h$ (in log scale) for a random 1-3-Horn-SAT formula where $d_3 = 3.0$ and $d_1 = d_1^*$. Let $d_1^*$ be the critical value of $d_1$ whose meaning and value are detailed in Appendix~\ref{app:tran}. $d_1^* \approx 0.098$.}
    \label{fig:critical}
    \Description{A standard strictly increasing line plot with just one line is drawn. Both X- and Y-axes are in log scale.}
\end{figure}

Figure~\ref{fig:critical} illustrates Observation~\ref{obs:3}, at one such point of discontinuity in the probability of satisfiability of $\mathcal{H}$, for $d_3 = 3.0$. The value of $d_1$ at which the discontinuity occurs is $d_1^* \approx 0.098$.

The three observations mentioned above empirically prove the Claim~\ref{cl:1}. \qed

\section{Conclusion}

In this paper, we provide strong empirical evidence to show that, for a random 1-3-Horn-SAT instance: (a) When $d_3 < 2$, the value of $h$ is proportional to  $\log n$ for all valid values of $d_1$, and (b) When $d_3 \ge 2$, the value of $h$ is proportional to  $\log n$ for all valid values of $d_1$ that exist outside of an $\epsilon$-neighborhood (for a small $\epsilon > 0$) of $d_1^*$, a value computed using Equation~\eqref{eq:d1*} in the appendices. Please note that within the $\epsilon$-neighborhood of $d_1^*$, the value of $h$ is still proportional to $\log n$ with the constant of proportionality increasing as $\epsilon$ gets smaller. At the exact point of singularity, i.e., $\epsilon = 0$, the value of $h$ is proportional to $n$.

This observation empirically supports that \textit{almost all} instances of a \textbf{P}-complete problem are in \textbf{NC}. This work also prompts programmers to look beyond worst-case inputs and encourages them to broaden their quest for efficient parallel algorithms for known hard problems, by figuring out how to contain pathological cases, where possible. 

%
\begin{acks}
We thank Moshe Vardi for suggesting the problem considered in this paper after seeing \citet{edwards2021study}. 
\end{acks}

\bibliographystyle{ACM-Reference-Format}
\bibliography{arXiv}

\newpage

\appendix

\section{Reduction from a general Horn formula to 3CNF Horn formula}\label{app:3cnf}
CNF Horn SAT, which allows an unlimited number of literals per clause up to one of them positive, has a simple reduction to 3CNF Horn-SAT, as demonstrated next. Consider the clause $(x_1 \vee \neg x_2 \vee \neg x_3 \vee \neg x_4)$. Replace it by the following two 3CNF Horn-SAT clauses $(x_1 \vee \neg x_2 \vee \neg x_5)$ and $(x_5 \vee \neg x_3 \vee \neg x_4)$, where $x_5$ is a new variable. 

If there are $k$ literals per CNF Horn clause ($k>3$), there are $k-3$ new variables added and the clause is split into $k-2$ 3CNF Horn clauses in the reduced 3CNF Horn formula.

\section{Description of the Horn-SAT input and output forms}\label{app:prob}
\subsection{Input form.} 
\begin{figure}[t]
     \centering
     \begin{subfigure}[b]{0.49\columnwidth}
         \centering
         \includegraphics[width=0.7\linewidth]{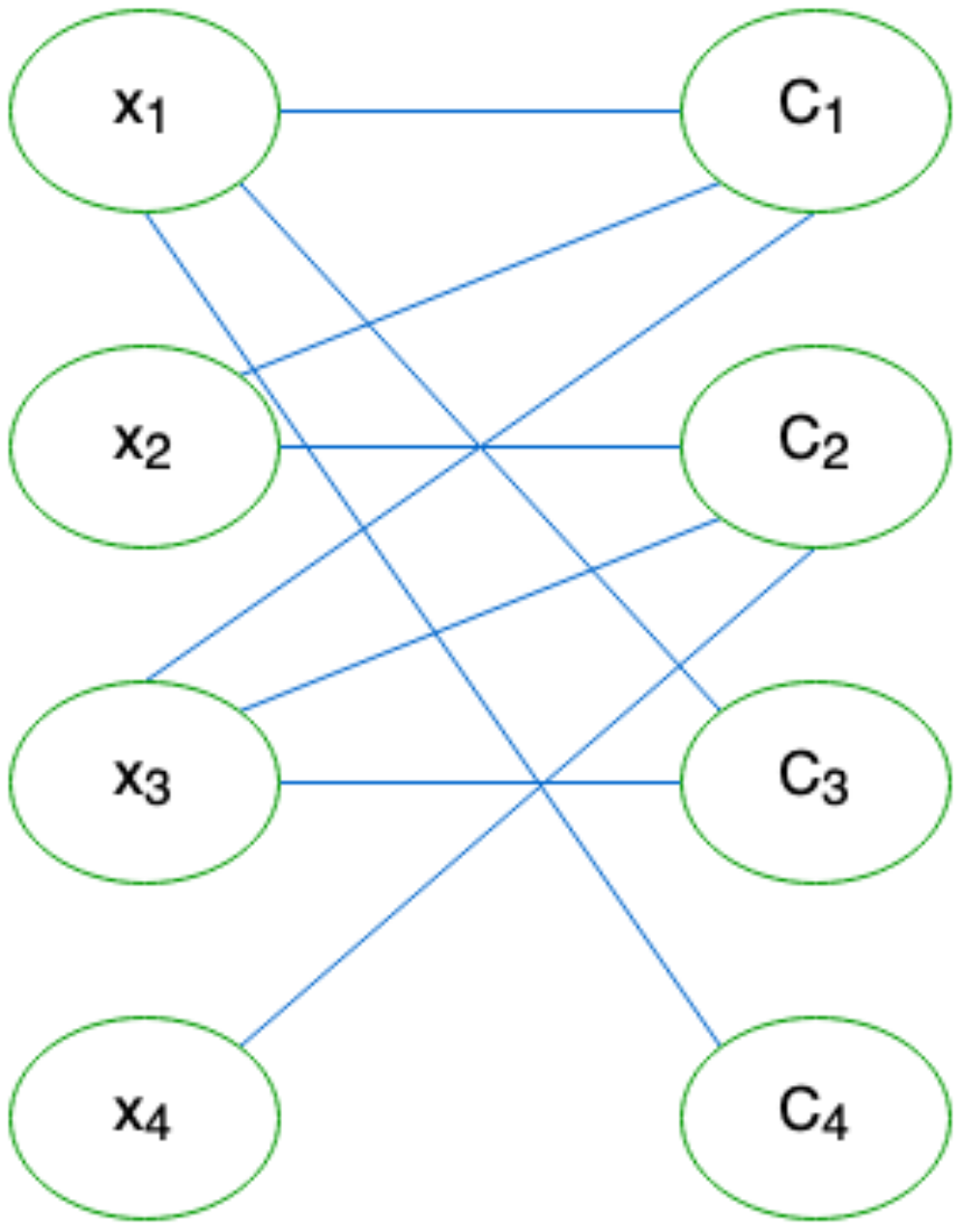}
         \caption{The bipartite graph for the 3CNF formula example.}
         \label{fig:1a}
     \end{subfigure}
     \hfill
     \begin{subfigure}[b]{0.49\columnwidth}
         \centering
         \includegraphics[width=0.8\linewidth]{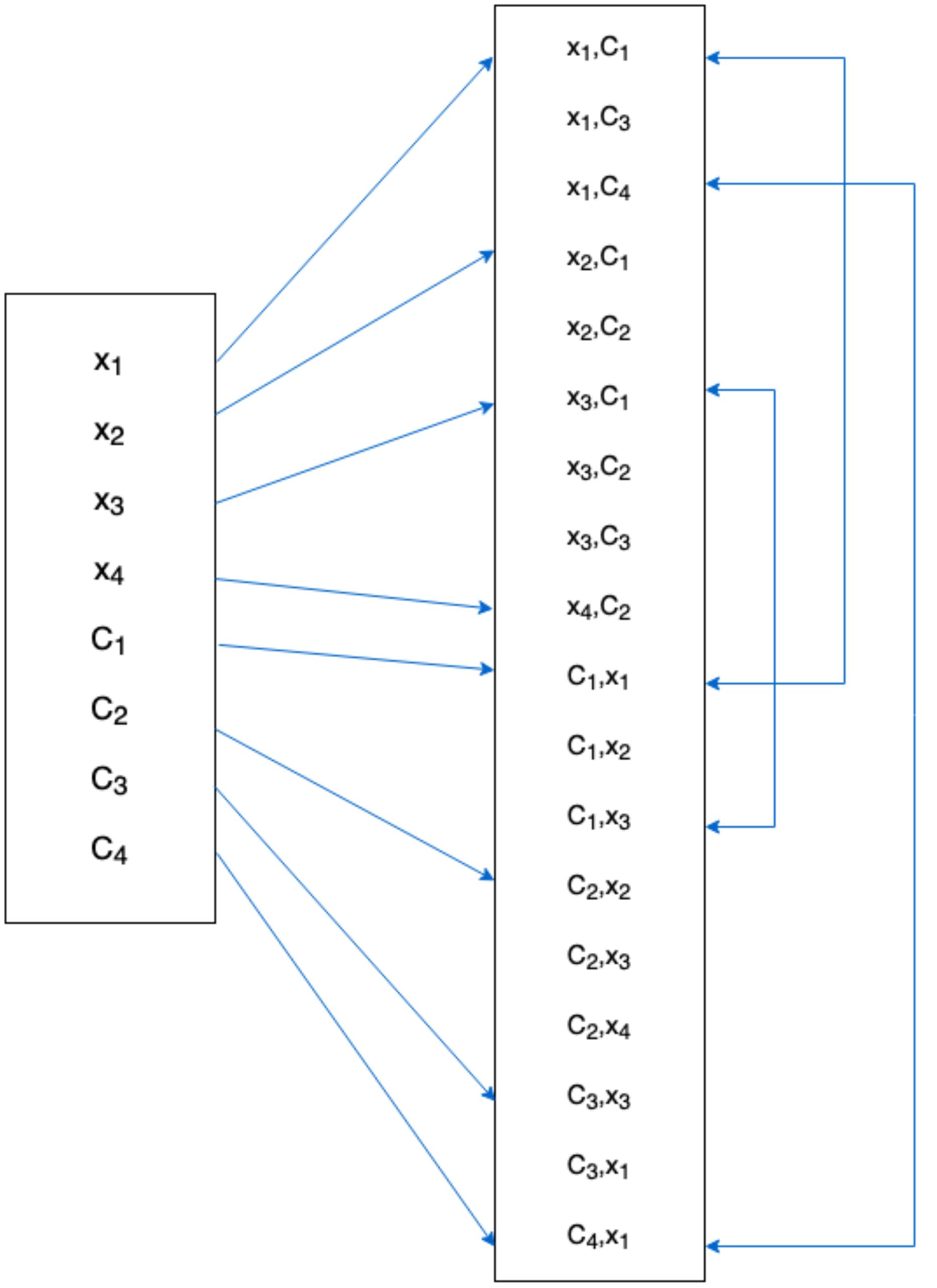}
         \caption{\textbf{Left column:} the vertex array. Variable vertices followed by the clause vertices. \textbf{Right column:} The edge array. Each variable vertex has a pointer to the beginning of the subarray of its literals. Each clause vertex has a pointer to the beginning of its literals subarray. Each edge in the bipartite graph appears twice: once for each of its vertices. Finally, each edge has a pointer to its other copy. Note: only an arbitrary subset of these pointers is shown.}
         \label{fig:1b}
     \end{subfigure}
        \caption{Input representation of the example 3CNF Horn-SAT formula, as used by the parallel algorithms described in this article. The Horn formula used in the figure is a conjunction of these clauses: $C_1 = \{x_1 \vee \neg x_2 \vee \neg x_3\}$, $C_2 = \{x_2 \vee \neg x_3 \vee \neg x_4\}$, $C_3 = \{x_3 \vee \neg x_1\}$, $C_4 = \{x_1\}$.}
        \Description{The left subfigure shows a bipartite graph with the left side of the graph comprising of variables and the right side of clauses. Undirected edges are drawn between clauses and variables contained in them in the formula. The right figure is the actual input form assumed by the algorithms presented in this article. There is a box with all variables and clauses listed in lexicographic order. There is another box with all combinations of related variables and clauses. There are directed edges between the two boxes and also among elements of the right box.}
        \label{fig:1}
\end{figure}

Array of size $n$ for variables $P_1, \ldots, P_n$. Array of size $m$ for clauses $C_1, \ldots, C_m$. Each clause has a link to a subarray of size at most 3 comprising its literals. Each variable has a link to a subarray of its literals in the clauses. This is presented in Figure \ref{fig:1} as a bipartite graph with $n+m$ nodes, one per variable, and one per clause. Each literal-clause pair provide an edge. We follow the standard input representation for parallel graph algorithms, such as in page 81 of \citet{vishkin2010thinking}.

\subsection{Output form.} A single boolean value indicating if the Horn-SAT instance is satisfiable or not. A Horn formula (or any logic formula) is satisfiable if there is a truth assignment to all of its variables such that the formula evaluates to \true{}. The formula is unsatisfiable if no such satisfying truth assignment exists. If the formula is satisfiable, give a satisfying truth assignment to all variables.

\subsection{Resolution.} All algorithms presented in this paper make use of ``resolution iterations'' to solve Horn formulae. A \textit{resolution} rule is an inference rule that produces a new clause implied by two clauses containing complementary literals. For example, consider a CNF formula $(\neg s \vee p) \wedge (q \vee \neg r \vee \neg p)$. Using the resolution rule, we can simplify the formula to $\neg s \vee q \vee \neg r$, removing the complementary literals of the variable $p$. 

\section{The Greedy parallel (GP) algorithm}\label{app:gp}
The pseudocode for the work-optimal parallel algorithm to solve Horn-SAT is given in Figure~\ref{algo:main}.

\begin{figure*}[ht]
\setlength{\fboxsep}{0pt}%
\setlength{\fboxrule}{0pt}%
\begin{center}
\begin{lstlisting}
While there are clauses with singleton literals ("unit clauses"):
  Satisfy all of these literals, do {
    In parallel, "Remove": 
      (i) all satisfied clauses, and 
      (ii) all implied FALSE literals in every clause in which they appear. For every affected clause, perform the following case analysis: 
        a. No literals remain. Conclude: the formula is unsatisfiable. Terminate the algorithm.
        b. Only one literal remained.  Include this literal in the set of singleton literals for the next iteration.
Optional step. If the current set of singleton literals does not include any positive literal, conclude: the formula is satisfiable. Assign FALSE to all remaining variables to derive a satisfying assignment. Terminate the algorithm. }
Final step. If the above did not already conclude that the formula is satisfiable or unsatisfiable, conclude: The formula is satisfiable. Assign FALSE to all remaining variables to derive a satisfying assignment.
\end{lstlisting}
\end{center}
\caption{The loop of the greedy parallel (GP) algorithm (parallel unit propagation)}
\label{algo:main}
\end{figure*}

\section{The Parallel Positive Unit Resolution (PPUR) Algorithm}\label{app:ppur}
 The PPUR algorithm, as stated in the main text, makes use of only resolution through positive unit clauses. Therefore, it can be seen as the parallelized version of the PUR algorithm for Horn-SAT, as presented in \citet{moore2007continuous}. The full algorithm is given in Figure~\ref{algo:ppur} for the sake of completeness. 

\begin{figure*}[ht]
\setlength{\fboxsep}{0pt}%
\setlength{\fboxrule}{0pt}%
\begin{center}
\begin{lstlisting}
While there are positive unit clauses:
  Satisfy all of these literals, do {
    In parallel, "Remove": 
      (i) all satisfied clauses, and 
      (ii) all implied FALSE literals in every clause in which they appear. For every affected clause, perform the following case analysis: 
        a. No literals remain. Conclude: the formula is unsatisfiable. Terminate the algorithm.
        b. Only one positive literal remains.  Include this literal in the set of positive singleton literals for the next iteration.
}
Final step. If the above did not already conclude that the formula is satisfiable or unsatisfiable, conclude: The formula is satisfiable. Assign FALSE to all remaining variables to derive a satisfying assignment.
\end{lstlisting}
    \end{center}
    \caption{The loop of the PPUR algorithm}
    \label{algo:ppur}
\end{figure*}
 
 The performance of the PPUR algorithm is upper bounded by the GP algorithm, but the main benefit of its presentation is that the different formulation allows our analysis to go through. 

\subsection{Depth of PPUR vs the GP algorithm} 
If a Horn-SAT instance is satisfiable, both of the PPUR and GP algorithms have the same depth, provided the optional step is used in the latter. In both cases, the algorithm iterates until no more positive unit clauses are generated and the rate of generating positive unit clauses is the same. The reason for the same rate is that negative unit resolution on Horn-SAT cannot generate positive unit clauses. 

If a Horn-SAT instance is unsatisfiable, the PPUR algorithm takes \textit{at most} twice the number of iterations taken by the GP algorithm. The unsatisfiablity will be concluded through unit resolution resulting in opposite truth assignment to the same variable, or if an empty clause is produced. We give an informal proof to the upper bound on depth of the PPUR algorithm:

Consider this pathological case for a parallel algorithm for Horn-SAT: Let the formula be $\mathcal{F} = \neg x_n \wedge x_1 \wedge (x_2 \vee \neg x_1) \wedge (x_3 \vee \neg x_2) \wedge (x_4 \vee \neg x_3) \dots \wedge (x_n \vee \neg x_{n-1})$. PPUR algorithm resolves two clauses (producing a unit clause) in every iteration and reaches a contradiction in $n$ iterations. On the other hand, the GP algorithm resolves four clauses in an iteration, reaching the contradiction in $n/2$ iterations. This is because the GP algorithm propagates negative unit clauses too, thereby attacking the formula from both ``ends'' of the resolution chain. 

We claim that no Horn-SAT instance exists that can be solved by the GP algorithm in faster than half the number of rounds it takes the PPUR algorithm to terminate. 

To prove the above claim, consider a Horn formula ($\mathcal{H}$) that is unsatisfiable on which the GP algorithm terminates in $k$ iterations. Assume the only initial negative unit clause is $\neg x_1$. It is implied that a positive and a negative unit clause of the same variable (say, $x_i$) is generated at the end of iteration $k$. At the end of iteration $k$ of the PPUR algorithm on the same formula $\mathcal{H}$, only the positive unit clause of the variable $x_i$ is generated. The algorithm now takes additional $k$ rounds to go through positive unit resolution to arrive at a contradiction with the variable $x_1$.


\section{Point of discontinuity of the satisfiability probability of random 1-3-Horn-SAT}\label{app:tran}
Here we give the results derived by \citet{moore2007continuous} for computing the points of discontinuity in probability of satisfiability (for any $d_3 \ge 2$), i.e., the points at which the formula switches from being highly likely to be satisfiable to highly likely to be unsatisfiable. 

Let $t_0 = \frac{1}{2}\left(1-\sqrt{1-\frac{2}{d_3}}\right)$, then the value of $d_1$ for which there is a phase transition in the satisfiability probability is given by:
\begin{equation}\label{eq:d1*}
  d_1^* = 1 - \frac{\mathrm{e}^{d_3t_0^2}}{2d_3t_0}
\end{equation}
Note that the discontinuity exists only for $d_3 \ge 2$. 

\section{Computing \texorpdfstring{$h$}{} for a run of the PPUR algorithm on a random 1-3-Horn-SAT formula}\label{app:h}

\citet{moore2007continuous} used Wormald's theorem \citep{wormald} to analyze the number of clauses of different lengths at each round of the serial PUR algorithm on a random Horn formula. 

\subsection{Recap of evolution of number of clauses of a random Horn formula in the serial algorithm}
Let the number of distinct clauses of length $j$ after $T$ stages of the PUR algorithm on a random Horn formula be denoted by $\mathbf{S_j}(T)$. Let $t = \frac{T}{n}$ be the scaled stage number (or, \textit{time}), normalized with respect to the number of variables $n$; $\mathbf{s_j}(t) = \frac{\mathbf{S_j}(T)}{n}$.

These $s_j$'s are scaled with respect to the original number of variables $n$, and $n$ at time $t$ is $n(t) = n(1-t)$. If one were to scale $S_j(t)$'s by $n(t)$, we obtain $\mathbf{d_j}(t) \coloneqq \mathbf{s_j}(t)/(1-t)$, for all $j \in \{1,2,3\}$. Based on the expressions of $\mathbf{s_j}(t)$'s given in Equation 3.4 of \citet{moore2007continuous}, the expressions for $\mathbf{d_j}(t)$'s are:

\begin{equation}\label{eq:serial}
\begin{aligned}
  \mathbf{d_1}(t) &= 1 - \frac{1-d_1}{1-t}\exp{-(d_2t + d_3t^2)}\\
  \mathbf{d_2}(t) &= (1-t) (d_2 + 2d_1d_3)\\
  \mathbf{d_3}(t) &= (1-t)^2d_3
\end{aligned}
\end{equation}

Note that $d_2 = 0$ for the model that is being studied.

\subsection{Evolution of number of clauses of a random Horn formula under the PPUR algorithm}
It can be seen that one round of the PPUR algorithm amounts to running the PUR algorithm on a random Horn formula until time $t=d_1$. This gives rise to the following recursive formulation of $d_j$'s for a run of the PPUR algorithm ($d_j^i$ denotes the value of $d_j$ at the beginning of round $i$):

\begin{equation}\label{eq:par}
\begin{aligned}
    n^{i+1} &\leftarrow n^{i}(1-d_1^i)\\
    d_1^{i+1} &\leftarrow 1-\exp{(-d_1^i(d_2^i + d_1^id_3^i))}\\
    d_2^{i+1} &\leftarrow (1-d_1^i)(d_2^i + 2d_1^id_3^i)\\
    d_3^{i+1} &\leftarrow d_3^i(1-d_1^i)^2
\end{aligned}
\end{equation}

\paragraph{Initial condition}
\begin{equation*}
    n^{0} \leftarrow n;\quad d_1^{0} \leftarrow d_1;\quad d_2^{0} \leftarrow 0;\quad d_3^{0} \leftarrow d_3
\end{equation*}

\paragraph{Termination condition}
The recursion terminates at iteration $h$ when the number of positive unit clauses generated is less than 1, i.e., $d_1^hn^h < 1$. Ideally, the value should be 0, but due to precision limitations on computers, we only require in this analysis the number of positive unit clauses to dip below $1$ to consider the algorithm to have converged.

\paragraph{Empirical validation}
The logarithmic value of $h$ as demonstrated in the direct simulation of the algorithm (as reported in \citet{vishkin2022beyond}) is in line with the logarithmic number of rounds implied by the above termination condition.

\end{document}